# Fractional Bi-Spectrum

Mehrdad Abolbashari, Gelareh Babaie, Jonathan Babaie, and Faramarz Farahi

*Abstract*—A signal with discrete frequency components, has a zero bispectrum if no linear combination of the frequencies equals one of the frequency components. We introduce fractional bispectrum in which for such signals the fractional bispectrum is nonzero. It is shown that fractional bispectrum has the same property as bispectrum for Gaussian signals: the fractional bispectrum of a zero mean Gaussian signal is zero; therefore it can be used to eliminate or reduce the Gaussian noise.

## I. INTRODUCTION

THE bispectrum is defined as the Fourier transform of the third order cumulant or moment of a stationary signal[1].

The bispectrum has some useful property. One of these properties is that the bispectrum of a stationary Gaussian signal with zero mean is zero; therefore if a signal is contaminated with a Gaussian noise, the bispectrum of the signal removes or reduces the effect of the noise. More precisely, if the signal is zero-mean and is independent of zero-mean Gaussian noise, then the effect of noise will be eliminated. Interestingly, a signal can be reconstructed uniquely from its bispectrum [1]; although there are some exception for the unique reconstruction of the signal. For example the bispectrum is insensitive to shift; i.e. the bispectrum of a signal is the same of the bispectrum of the shifted version of the same signal. Therefore it is possible to remove or reduce the amount of the noise using the bispectrum.

Another property of the bispectrum is preservation of the phase; bispectrum can be used to extract the phase information [2], [3] which is lost in Power Spectrum. In addition, the bispectrum can be used to reconstruct the amplitude of the Fourier transform of the signal [4]; therefore the signal can be reconstructed from its bispectrum.

One application of the bispectrum is for system identification [5] by computing the impulse response of an LTI system using bispectrum. Speckle noise reduction is another application for bispectrum [6]; the multiplicative speckle noise first is converted to additive noise using logarithm and by assuming that the logarithm of speckle noise is estimated with an independent Gaussian noise, the speckle noise is reduced.

For a stationary signal with zero mean the third order moment and cumulant are the same and are defined as

$$C(\rho,\tau) = E\{x(t+\rho)x(t+\tau)x(t)\} \quad (1)$$

where $E\{.\}$ shows the expected value.
The Fourier transform of cumulant is defined as bispectrum

$$S(u,v) = \mathfrak{F}\{C(\rho,\tau)\} = \mathfrak{F}\{E\{x(t+\rho)x(t+\tau)x(t)\}\} \quad (2)$$

where $\mathfrak{F}\{.\}$ shows the Fourier transform.
We adopt the definition of Lohmann and Wirnitzer [2] for the cumulant as follows

$$R(\rho,\tau) = \int_{-\infty}^{+\infty} x(t+\rho)x(t+\tau)x(t)dt \quad (3)$$

Where $x(t)$ is an instance of the stochastic process $\mathbf{x}(t)$ and $R(\rho,\tau)$ is an estimation of the cumulant of $\mathbf{x}(t)$.
Therefore the bispectrum can be calculated as

$$\begin{aligned}
S(u,v) = \mathfrak{F}\{R(\rho,\tau)\} &= \iint_{-\infty}^{+\infty} R(\rho,\tau)e^{-i2\pi(u\rho+v\tau)}d\rho d\tau \\
&= \iiint_{-\infty}^{+\infty} x(t+\rho) \\
&\quad x(t+\tau)x(t)e^{-i2\pi(u\rho+v\tau)}dt d\rho d\tau \\
&= \int_{-\infty}^{+\infty} \left( \int_{-\infty}^{+\infty} x(t+\rho)e^{-i2\pi u(t+\rho)}d\rho \right) \\
&\quad \left( \int_{-\infty}^{+\infty} x(t+\tau)e^{-i2\pi v(t+\tau)}d\tau \right) x(t)e^{-i2\pi t(-u-v)}dt \\
&= X(u)X(v)X(-u-v) \quad (4)
\end{aligned}$$

where $X(.)$ is the Fourier transform of the signal $x(.)$.
For a real signal $X(-u) = X^*(u)$ and therefore the bispectrum can be written as

$$S(u,v) = X(u)X(v)X(-u-v) = X(u)X(v)X^*(u+v) \quad (5)$$

As it is seen in (5), in order for bispectrum to be nonzero there should be nonzero components at frequency $u = f_1$,

Mehrdad Abolbashari is with the OPTONIKS CORP., PORTAL Building, 9319 Robert D. Snyder Road, Charlotte, NC 28223, USA (e-mail: m.abolbashari@optoniks.com).
Gelareh Babaie (e-mail: gbabaie@uncc.edu), Jonathan Babaie (e-mail: jbabaie@uncc.edu) and Faramarz Farahi (e-mail: ffarahi @uncc.edu) are with The Center for Optoelectronics and Optical Communications and The Center for Precision Metrology, University of North Carolina at Charlotte, Charlotte, NC 28223, USA.

---

[1] The definition is given as both the Fourier transform of the moment [7] or cumulant [8]. For a stationary stochastic signal with zero mean, the moment and cumulant are equal.

$v = f_2$ and $u + v = f_1 + f_2$ for some frequencies. If such combination did not exist the bispectrum of the signal would be zero. For example, the bispectrum of a pure sinusoidal signal is zero. Also the bispectrum of the combination of two sinusoidal signals is zero if the frequency of one signal is not twice as the frequency of the other one.

Obviously, in most cases the signals have zero bispectrum and the benefits of the bispectrum cannot be sought. It is in fact desirable to have nonzero bispectrum even if the frequencies do not satisfy the required relation. This happens, for example, where the signal is the combination of two laser sources in which the wavelength of one of the laser sources is not exactly twice as the other's form a new signal (which is very probable). In this letter we describe a technique that expands the application of the bispectrum.

We introduce fractional bispectrum and show that the function holds the very attractive property of bispectrum since for a zero mean Gaussian signal, the fractional bispectrum is zero. We define the fractional bispectrum as

$$F(u, v, k) = X(u)X(v)X(-u - kv)$$
$$= X(u)X(v)X^*(u + kv) \quad (6)$$

where $k \in \mathbb{R}$. As it is seen, the bispectrum is a special case of fractional bispectrum where $k = 1$.

As it was mentioned above, the bispectrum of a Gaussian signal is zero; therefore if a signal has additive Gaussian noise, by using the bispectrum, the effect of Gaussian noise will be removed or reduced. We will show below that this is true for fractional bispectrum, too. To show that the fractional bispectrum of a Gaussian signal with zero mean is zero, it suffices to show that the inverse Fourier transform of the fractional bispectrum is a third order statistics, and since the third and higher order statistics of a Gaussian signal is zero, therefore for the Gaussian signal with zero mean, the fractional bispectrum will be zero.

The inverse Fourier transform of fractional bispectrum is calculated as

$$R_F(\rho, \tau, k) = \mathfrak{F}^{-1}\{F(u, v, k)\}$$
$$= \iint_{-\infty}^{+\infty} F(u, v, k) e^{i2\pi(u\rho + v\tau)} du\, dv$$
$$= \iint_{-\infty}^{+\infty} X(u)X(v)X(-u - kv) e^{i2\pi(u\rho + v\tau)} du\, dv$$
$$= \iint_{-\infty}^{+\infty} \int_{-\infty}^{+\infty} x(t_1) e^{-i2\pi u t_1} dt_1 \int_{-\infty}^{+\infty} x(t_2) e^{-i2\pi v t_2} dt_2$$
$$\int_{-\infty}^{+\infty} x(t_3) e^{-i2\pi(-u-kv)t_3} dt_3 \, e^{i2\pi(u\rho + v\tau)} du\, dv$$
$$= \iiint_{-\infty}^{+\infty} x(t_1)x(t_2)x(t_3) \int_{-\infty}^{+\infty} e^{i2\pi u(\rho + t_3 - t_1)} du$$
$$\int_{-\infty}^{+\infty} e^{i2\pi v(\tau + kt_3 - t_2)} dv \, dt_1 dt_2 dt_3 \quad (7)$$

$$= \iiint_{-\infty}^{+\infty} x(t_1)x(t_2)x(t_3) \delta(\rho + t_3 - t_1)$$
$$\delta(\tau + kt_3 - t_2) dt_1 dt_2 dt_3$$
$$= \int_{-\infty}^{+\infty} x(\rho + t_3)x(\tau + kt_3)x(t_3) dt_3$$

Therefore

$$R_F(\rho, \tau, k) = \int_{-\infty}^{+\infty} x(\rho + t)x(\tau + kt)x(t) dt$$
$$= E\{x(\rho + t)x(\tau + kt)x(t)\} \quad (8)$$

Since $\rho$, $\tau$, and $t$ are independent variables; the $R_F(\rho, \tau, k)$ is a scaled and shifted function of $R(\rho, \tau)$; and since for a Gaussian signal with zero mean $R(\rho, \tau) = 0$, therefore for the Gaussian signal with zero mean $R_F(\rho, \tau, k) = 0$, and as a result $F(u, v, k) = \mathfrak{F}\{R_F(\rho, \tau)\} = 0$.

It is worth to mention that in practice, since the signal $x(t)$ is one instance of a zero mean Gaussian stochastic process and the $R(\rho, \tau)$ is an estimation of the cumulant, the fractional bispectrum of a signal $x(t)$ might not be zero, but would be a small value and as a result by using the fractional bispectrum function, the zero mean Gaussian noise would be reduced instead of being eliminated.

To summarize, we have introduced the fractional bispectrum transformation, as a generalization of bispectrum transformation. The fractional bispectrum is useful in situations that the bispectrum of the signal is zero, while there are some set of nonzero frequency components of $\{u, v, u + kv\}$ for some $k \in \mathbb{R}$. This situation happens, for example, when the signal comprised of two sinusoidal signals where the frequency of one sinusoidal signal is not twice of other. Another example is where the signal is a bandpass signal and the highest nonzero frequency component is smaller than the twice of the lowest nonzero frequency component. In both cases the bispectrum of the signal is zero, while the fractional bispectrum would be nonzero by choosing the proper value for $k$. Fractional bispectrum has the advantages offered by the bispectrum transformation but strict limitation related to bispectrum is no longer applied.